\documentclass[a4paper,fleqn,usenatbib]{mnras}

\usepackage{mathptmx}

\usepackage[T1]{fontenc}
\usepackage{ae,aecompl}

\usepackage{graphicx}  
\usepackage{amsmath}   
\usepackage{amssymb}   

\newcommand{\gccm}{\,g cm$^{-3}$}
\newcommand{\Kccm}{\,K cm$^{-3}$}

\title[Non-linear core formation in L1517]{Non-linear dense core formation in the dark cloud L1517}

\author[S. Heigl et al.]{
S. Heigl,$^{1,2}$\thanks{E-mail: heigl@usm.lmu.de}
A. Burkert,$^{1,2}$
A. Hacar,$^{3}$
\\
$^{1}$Universit\"ats-Sternwarte M\"unchen, LMU Munich, Scheinerstr. 1, 81679 Munich, Germany\\
$^{2}$Max-Planck Institute for Extraterrestrial Physics, Giessenbachstr. 1, 85748 Garching, Germany\\
$^{3}$Institute for Astrophysics, University of Vienna, T\"urkenschanzstrasse 17, 1180 Vienna, Austria
}

\date{Accepted XXX. Received YYY; in original form ZZZ}

\pubyear{2016}

\begin{document}
\label{firstpage}
\pagerange{\pageref{firstpage}--\pageref{lastpage}}
\maketitle

\begin{abstract}
   We present a solution for the observed core fragmentation of filaments in
   the Taurus L1517 dark cloud which previously could not be explained
   \citep{hacar2011}.
   Core fragmentation is a vital step for the formation of stars.
   Observations suggest a connection to the filamentary structure
   of the cloud gas, but it remains unclear which process is responsible.
   We show that the gravitational instability process of an infinite, isothermal
   cylinder can account for the exhibited fragmentation under the assumption
   that the perturbation grows on the dominant wavelength.
   We use numerical simulations with the code RAMSES, estimate observed
   column densities and line-of-sight velocities, and compare them to the
   observations.
   A critical factor for the observed fragmentation is
   that cores grow by redistributing mass within the filament and thus
   the density between the cores decreases over the fragmentation
   process. This often leads to wrong dominant wavelength estimates, as it
   is strongly dependent on the initial central density.
   We argue that non-linear effects also play an important role on
   the evolution of the fragmentation.
   Once the density perturbation grows above the critical line-mass,
   non-linearity leads to an enhancement of the central core density in
   comparison to the analytical prediction.
   Choosing the correct initial conditions with perturbation strengths
   of around 20\%, leads to inclination corrected line-of-sight velocities
   and central core densities within the observational measurement error
   in a realistic evolution time.
\end{abstract}




\section{Introduction}
\label{sec:introduction}

   Individual or binary stars are formed in dense cores
   \citep{benson1989,difrancesco2007} which are condensations within larger
   molecular cloud complexes.
   The critical process of how a tens of parsec sized cloud fragments
   into a few $0.1 \text{ pc}$ sized cores nevertheless remains an unresolved
   challenge for star formation. Especially, as core formation appears to be
   tied to the low efficiency of the star-formation process \citep{evans2009}.

   The recent large-scale cloud images taken by the \textit{Herschel} Space
   Observatory show that molecular clouds exhibit a ubiquity of complex
   filamentary structures, forming a network over several size scales
   \citep{andre2010,molinari2010,arzoumanian2011,schneider2012} and dense 
   cores being aligned with
   large-scale filaments like pearls in a string. This is a strong
   indication that core formation is tied to some kind of
   filament fragmentation process, a connection which has long been proposed
   \citep{schneider1979,larson1985}.

   Recent observations however have shown that filamentary clouds with
   trans-sonic internal motions are not one single entity, but consist of
   fibers: velocity-coherent structures of subsonic gas
   \citep{hacar2013, arzoumanian2013, tanaka2013, tafalla2015}. The existence
   of filamentary substructure has also been found recently in numerical
   simulations \citep{moeckel2015, smith2016}. Dense cores are embedded within
   these fibers and often show similar kinematic properties with a smooth
   transition from fiber to core gas. This suggests that turbulence does not
   dissipate at the scale of dense cores, but at the typical scale of the
   velocity-coherent fibers of about $0.5 \text{ pc}$. \citet{tafalla2015}
   propose a model they call "\textit{fray and fragment}":
   At first, the filament forms through a colliding flow. Over time, residual
   turbulent motions together with gravity form velocity-coherent fibers.
   Finally, some of the fibers form dense cores through gravitational
   instability. The model suggests that cores form by subsonic motions and gives
   rise to the question of the exact mechanism that leads to core formation in
   fibres and filaments. The key to understanding the core fragmentation process
   and to distinguish between different models, such as a pure gravitational
   fragmentation, dissipation of turbulence due to supersonic shocks
   \citep{padoan2001, klessen2005, vazquez-semadeni2005} or the loss of magnetic
   support due to ambipolar diffusion \citep{shu1987}, is the internal
   velocity structure of the filaments, as well as their kinematic properties.

   A detailed study of the L1517 dark cloud in Taurus and its core population
   was presented by \citet{hacar2011}. They observed the cloud in four
   different molecular transitions, ranging from
   $\text{N}_2\text{H}^+(J=1-0)$ and SO$(J_N=3_2-2_1)$ to
   $\text{C}^{18}\text{O}(J=1-0)$ and $\text{C}^{17}\text{O}(J=1-0)$
   as well as in dust continuum emission.
   This allows them to study different density regimes, from the less dense
   filament gas to the very dense core interior, in great detail. In addition
   they show that the cores are in different evolutionary states, with more
   evolved cores having increasing N$_2$H$^+$ abundance and anti-correlated,
   depleted SO emission \citep{tafalla2006}. The region is made up of four
   velocity coherent filaments consisting predominantly of subsonic gas.
   In two of the filaments, two cores are forming separately and the
   interior motion of the filaments show a smooth transition to the core gas
   kinematics. The line-of sight velocity centroid shows an oscillatory motion
   with a periodicity that matches the periodicity of the cores and is clearly
   associated with the core positions. This oscillatory motion is expected from the
   gravitational instability model and suggests that the cores form by gravitational
   contraction of the filament gas. Nevertheless, the authors claim that the
   fragmentation distance is not consistent with the model of pure gravitational
   fragmentation, a claim we want to test using numerical methods as the model
   is strongly dependent on its initial condition and the inclination of the filament.

   In the following sections, we recapitulate the filament and core
   population of L1517 (\autoref{sec:L1517population}) and the characteristic
   properties our models have to replicate. We also summarize the theory of the
   fragmentation of pressure-bound isothermal filaments (\autoref{sec:theory}),
   the basis of our models and how we use it in our simulations
   (\autoref{sec:prediction}). Next, we discuss the simulation set-up
   (\autoref{sec:simulationsetup}) and present our numerical results
   (\autoref{sec:simulations}). We start by comparing our data to the
   observations (\autoref{sec:observations}), then we discuss the evolution of
   the core growth in the linear and non-linear regime (\autoref{sec:evolution}).
   Finally, we finish with an analysis of the external pressure in the dark cloud
   (\autoref{sec:externalpressure}) which is a crucial part of the model of a
   pressure-bound isothermal filament.

\section{Filaments and cores in L1517}
\label{sec:L1517population}

   The L1517 dark cloud has several properties in favour of an idealized
   analysis. The cloud is relatively isolated, only to be disturbed
   by the near PMS stars AB Aurigae and SU Aurigae, which are
   physically associated with L1517 and are heating the near-by gas
   \citep{nachman1979,duvert1986,heyer1987}, although no influence on the
   dense gas can be seen. Furthermore, the filament gas is predominantly
   subsonic. Thus, turbulence only plays a minor role in the dynamics of
   the gas.

   The density profiles of three of the four
   filaments can be reproduced by an isothermal cylinder in
   pressure equilibrium with its self-gravity. This profile was first
   described by \citet{stodolkiewicz1963} and \citet{ostriker1964}, and is
   given by the analytic form
   \begin{equation}
      \rho(r) = \frac{\rho_c}{\left(1+\left(r/H\right)^{2}\right)^{2}}
      \label{eq:radeq}
   \end{equation}
   where $r$ is the cylindrical radius, $\rho_0$ is the central density. The
   radial scale height $H$ is given by
   \begin{equation}
      H^2 = \frac{2c_s^2}{\pi G \rho_c}
      \label{eq:scaleheight}
   \end{equation}
   where $c_s$ is the isothermal sound speed and $G$ the gravitational
   constant. The gas is assumed to have a temperature of approximately
   $10 \text{ K}$ \citep{tafalla2004} and assuming a molecular weight of
   $\mu=2.36$ gives the isothermal sound speed of
   $c_s\approx0.2 \text{ km s}^{-1}$. Integrating the profile to
   $r \rightarrow \infty$ gives a line mass of
   \begin{equation}
      \left(\frac{M}{L}\right)_\text{crit} =
      \frac{2c_s^2}{G}\approx16.4 \text{ M}_{\sun}\text{ pc}^{-1}
   \end{equation}
   This is called the critical line-mass, as it determines the threshold
   above which a filament will collapse under its self-gravity.

   It is important to note that the observations do not extend out far enough
   in radius to distinguish between different outer density profiles. The
   filaments could also be reproduced by a shallower softened power law and
   it is practically impossible to say if they are in pressure equilibrium.
   Although, the fact that they form well-separated cores \citep{inutsuka1997}
   and do not show supersonic motions \citep{burkert2004} are good indicators
   that they indeed are in pressure equilibrium.
   Only one of the filaments can be better reproduced by using a softened power
   law and does not follow the isothermal profile. We will concentrate on
   filaments 1 and 2, following the nomenclature of \citet{hacar2011}, both
   of which show an agreement with a profile in pressure equilibrium.
   They exhibit a prominent core fragmentation for which detailed measurements
   of densities and line-of-sight velocities along the filament are available.

   Filament 1 has a total mass of $M_\text{fil} = 8.0 \text{ M}_{\sun}$ and
   an observed projected length $L_\text{obs} = 0.52 \text{ pc}$. It contains
   the cores A2 and C with central number densities of
   $6.0 \cdot 10^4 \text{ cm}^{-3}$ and $4.7 \cdot 10^4 \text{ cm}^{-3}$
   which, assuming a molecular weight of $\mu = 2.36$, correspond to
   $2.4 \cdot 10^{-19} \text{\gccm}$ and
   $1.8 \cdot 10^{-19} \text{\gccm}$, respectively.
   The filament profile fit provides a central number density between the cores
   of about $10^4 \text{ cm}^{-3}$ or $3.9 \cdot 10^{-20} \text{\gccm}$.
   The observed projected core distance is about $340 \text{ arcsec}$ or
   $0.23 \text{ pc}$, assuming a distance of $144 \text{ pc}$ to AB Aur
   \citep{vandenancker1998}. The line-of-sight velocity centroid variation
   along the filament shows a core forming motion following a sinusoidal
   pattern with an amplitude of 0.04 km s$^{-1}$ after subtracting a smooth linear
   gradient of 1.0 km s$^{-1}$ pc$^{-1}$.

   Filament 2 is measured to have a total mass of
   $M_\text{fil} = 7.2 \text{ M}_{\sun}$ and an observed projected length
   $L_\text{obs} = 0.42 \text{ pc}$. It contains the cores A1 and B,
   which have a central density of $7.0 \cdot 10^4 \text{ cm}^{-3}$ and
   $2.2 \cdot 10^5$ cm$^{-3}$, which corresponds to
   $2.7 \cdot 10^{-19} \text{\gccm}$ and
   $8.6 \cdot 10^{-19} \text{\gccm}$ respectively. The central density
   between the cores is determined to be $7.0 \cdot 10^3 \text{ cm}^{-3}$
   which is  equivalent to $2.7 \cdot 10^{-20} \text{\gccm}$. The observed
   core distance is about $270\text{ arcsec}$ or $0.19 \text{ pc}$. In
   contrast to filament 1, the line-of-sight velocity centroid variation
   does not follow a well defined pattern as filament 1. Nevertheless,
   \citet{hacar2011} fit a linear gradient of 1.4 km s$^{-1}$ pc$^{-1}$ and
   a sinusoidal pattern with an amplitude of 0.04 km s$^{-1}$ to the data. While
   it does match the observed velocity pattern in certain regions, it fails
   to explain the overall form.

\section{Filament fragmentation}

\subsection{Theory of filament fragmentation}
\label{sec:theory}

   There has been extensive theoretical work on the fragmentation of infinite,
   isothermal filaments over the last fifty years. In reality, filaments are neither
   infinite nor isothermal. The approximation of isothermality is probably a valid
   approach in the case of L1517 as the density profiles match the isothermal profile.
   Also the typical dust temperature gradients in filaments are smaller than a
   few Kelvin \citep{arzoumanian2011, palmeirim2013}. The bigger caveat is that filaments
   have a finite length. It has already been shown that filaments collapse globally via the
   end-dominated mode where clumps form at both ends of the filament due to gravitational
   focusing \citep{bastien1983, burkert2004, pon2012}. But it still remains unclear how
   equilibrium filaments fragment exactly under global collapse. While solutions have been
   found for a radial equilibrium of non-isothermal filaments \citep{recchi2013}, they only
   apply to infinite filaments. This is also true for the above mentioned radial solution
   found by \citet{stodolkiewicz1963}. There is still a lack of detailed theoretical
   studies on the structure and fragmentation of finite filaments.
   \citet{bastien1991} looked at the fragmentation of finite cylinders with a uniform density
   profile. They discover a similar behaviour as for the pressure truncated infinite
   equilibrium case presented below and also find a critical wavelength beneath which
   density perturbations will not grow. It differs by a factor of four from the infinite
   filament case predicting more fragments in finite filaments. They also find that in most
   cases it is possible to form growing fragments along the cylinder before a complete
   collapse and that the dominance of the end fragments decreases for a higher mass of the
   clouds. But the study still misses a detailed numerical prediction of dominant
   fragmentation scales and its dependence on the line-mass. We therefore stress the fact
   that due to a lack of a better theory we use the approximation of an infinite,
   isothermal filament which we present here.
   \newline
   \newline
   We introduce a small density perturbation in the linear regime along the
   filament axis of the form:
   \begin{equation}
      \rho(r,z,t)=\rho_0(r)+\rho_1(r,z,t)=
                  \rho_0(r)+\epsilon\rho_0(r)\exp(ikz-i\omega t)
      \label{eq:rhoperturbation}
   \end{equation}
   In this case, $z$ is the filament axis, $\omega$ is the growth rate,
   $k=2\pi/\lambda$ the wave vector and $\epsilon$ the perturbation
   strength. Neglecting second order terms, perturbations will grow
   for values of $k$ where the solution of the dispersion relation
   $\omega^2(k)$ is smaller than zero. This will also lead to a perturbation
   in velocity, pressure and potential of the form:
   \begin{equation}
      q_1(r,z,t) \propto \exp(ikz-i\omega t)
      \label{eq:perturbation}
   \end{equation}
   It was shown that there are two important parameters for the fragmentation of
   an infinite, isothermal filament: the critical and the dominant wavelength.
   On the one hand, the critical wavelength determines the separation above
   which a small perturbation will grow. It was first determined by
   \citet{stodolkiewicz1963} to be $\lambda_\text{crit}=3.94H$ for
   a filament extending to infinite radius. The dominant wavelength, on the
   other hand, gives the separation of the perturbation which will grow
   the fastest. It is therefore the most likely perturbation length a filament
   will show after letting random perturbations grow. It was first
   determined by \citet{larson1985} to be about twice the critical
   wavelength: $\lambda_\text{dom}=7.82H$ with a growth rate of
   $\left|\omega_\text{dom}\right|=0.339\left(4\pi G\rho_c\right)^{1/2}$
   \citep{nagasawa1987, inutsuka1992, nakamura1993, gehman1996}.

   \citet{nagasawa1987} was the first to also consider the more realistic
   situation of a pressure truncated filament. In this case the filament
   follows the pressure equilibrium profile until it extends to the radius
   where the internal pressure matches the external pressure.
   The external pressure stabilizes the filament against expansion and
   filaments below the critical line-mass do not extend to infinity and
   are stable.
   The factor of line-mass to critical line-mass is given by
   \begin{equation}
      f_\text{cyl}=\left.\left(\frac{M}{L}\right)\middle/
      \left(\frac{M}{L}\right)_\text{crit}\right.
   \end{equation}
   This leads to the boundary radius of
   \begin{equation}
      R=H\left(\frac{f_\text{cyl}}{1-f_\text{cyl}}\right)^{1/2}
   \end{equation}
   and the boundary density of
   \begin{equation}
      \rho_b=\rho_0\left(1-f_\text{cyl}\right)^2
      \label{eq:rhoboundary}
   \end{equation}
   For $f_\text{cyl}\rightarrow1$ the dispersion relation tends to the same
   dispersion relation as for the non-truncated filament. In the case that the
   filament exceeds the critical line-mass $(f_\text{cyl} > 1)$,
   \citet{inutsuka1992} demonstrated that the filament collapses faster to the
   axis than perturbations can grow.

   \citet{nagasawa1987} chose an external density of zero for the computation
   of dominant length- and timescales, corresponding to an infinite temperature.
   But even for a non-infinite external temperature of ten times greater than
   the filament temperature, \citet{fiege+pudritz2000} found no difference
   to the case of an infinite external temperature. In the isothermal
   case this is equal to setting the external density to a value of ten
   times less than the boundary density. We also follow this approach
   and set the external density even lower in order to reduce the effect of
   accretion onto the filament. A more realistic approach would
   consider that observations show a smooth transition of filaments
   into the surrounding medium \citep{arzoumanian2011, palmeirim2013}.

   The findings on dominant fragmentation lengthscales and growth timescales
   for pressure truncated filaments by \citet{nagasawa1987} were summarized and
   interpolated by \citet{fischera+martin2012} using a fifth-order polynomial
   function with vanishing derivatives at the extremes:
   \begin{equation}
      y(x)=\sum_{i=0}^{5}a_if_\text{cyl}^{i/2}
   \end{equation}
   They find the interpolation values given in
   \autoref{table:constants}. As one can see the dominant timescale
   $\tau_{\text{dom}}$, the critical wavelength $\lambda_{\text{crit}}$ and the
   dominant wavelength $\lambda_{\text{dom}}$ are all polynominals of different
   power of $f_{cyl}$. FWHM is the approximated full width half maximum of the
   filament. \autoref{table:constants} makes it possible to express the
   dominant timescale and the dominant wavelength as a factor of the scale height
   $H$.
   \begin{table}
      \centering
      \caption{Constants of polynomial approximations by
      \citet{fischera+martin2012}. FWHM is the approximated
      full width half maximum of the filament. $\tau_\text{dom}=1/\omega_\text{dom}$
      is the growth timescale of the dominant mode given in units of
      $\sqrt{4\pi G \rho_c}$.}
      \label{table:constants}
      \resizebox{\hsize}{!}{
      \begin{tabular}{c c c c c c c}
         \hline\hline
         & $a_0$ & $a_1$ & $a_2$ & $a_3$ & $a_4$ & $a_5$ \\
         \hline
         $\tau_\text{dom}\sqrt{4\pi G \rho_c}$ & 4.08 & 0.00 & -2.99 & 1.46 & 0.40 & 0.00 \\
         $\lambda_\text{crit}/\text{FWHM}$ & 3.39 & 0.00 & -2.414 & 1.588 & 0.016 & 0.00 \\
         $\lambda_\text{dom}/\text{FWHM}$ & 6.25 & 0.00 & -6.89 & 9.18 & -3.44 & 0.00 \\
         $\text{FWHM}/H$ & 0.00 & 1.732 & 0.00 & -0.041 & 0.818 & -0.976 \\
         \hline
      \end{tabular}
      }
   \end{table}
   We also use this interpolation for our simulations to define our initial
   density perturbation length and to find the expected growth timescale of
   the dominant mode.

\subsection{Analytical prediction for L1517}
\label{sec:prediction}

   The linear model relies on an exponential growth of a sinusoidal density
   and velocity perturbation as seen from \autoref{eq:rhoperturbation}.
   This implies that the perturbation is symmetrical in the sense that as
   the density enhancement grows, the density minimum depletes on the same
   timescale. Thus, as soon as the peak density reaches twice the
   initial density, the gas between the cores should be completely accreted
   and the model has to break down. Therefore, it is not clear how long a
   sinusoidal redistribution of mass is maintained.

   For a first test, we adopt an initial central density that is the simple
   average between the filament and maximum core density, where we take the
   average of both cores respectively as density maximum. This leads to the
   values of about $1.2 \cdot 10^{-19} \text{\gccm}$ for filament 1 and about
   $2.9 \cdot 10^{-19} \text{\gccm}$ for filament 2, corresponding to the
   number densities of $3.1 \cdot 10^{4} \text{ cm}^{-3}$ and
   $7.4 \cdot 10^{4} \text{ cm}^{-3}$ respectively. In the case we do not
   achieve a minimum density as low as the observed minimum density when
   running the fragmentation simulation, we use the mean of our guess value
   and the observed minimum central density to calculate a new initial central
   density and iterate until a good agreement with the observation is found.

   We expect the cores to grow on the dominant wavelength and thus the
   observed core separation is the dominant wavelength affected by
   inclination. We use the inclination angle $\phi$ where the projected
   dominant wavelength $\lambda_\text{dom}\cdot\cos(\phi)$
   corresponds to the observed fragmentation length. The value of
   $\lambda_\text{dom}$ is computed using \autoref{table:constants},
   where $H$ is calculated according to \autoref{eq:scaleheight} and the
   line-mass fraction $f_\text{cyl}$ is determined by using the filament
   mass M$_\text{fil}$ and the observed filament length $L_\text{obs}$,
   which is corrected for inclination:
   \begin{equation}
      f_\text{cyl} = \frac{M_\text{fil}\cdot\cos(\phi)}{L_\text{obs}}\left/
      \left(16.4 \text{ M}_{\sun}\text{ pc}^{-1}\right)\right.
   \end{equation}
   With the central density and $f_{cyl}$ determined, we have
   everything we need to set-up a filament in pressure equilibrium.

\section{Simulation set-up}
\label{sec:simulationsetup}

   \begin{figure*}
      \includegraphics[width=\textwidth]{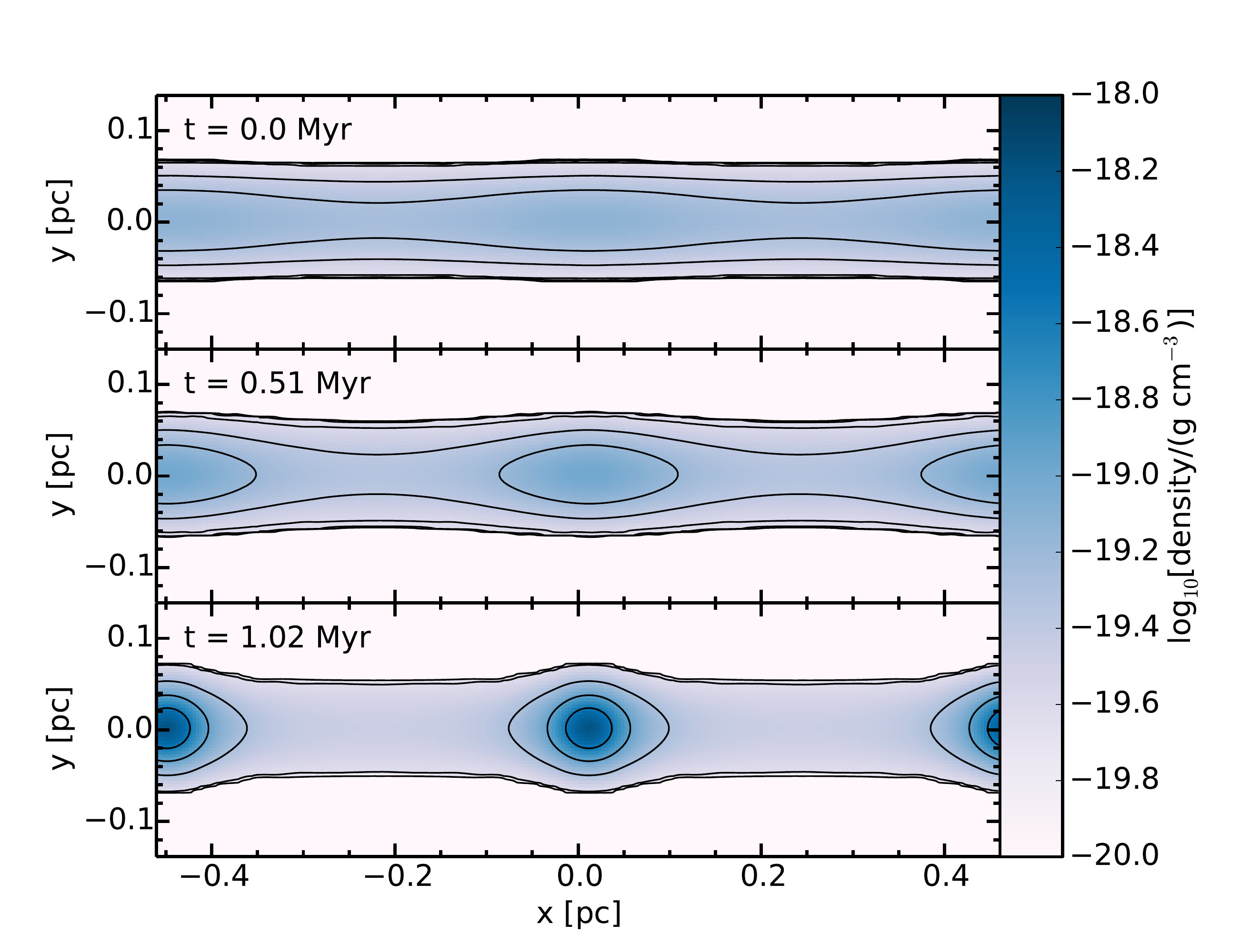}
      \caption{Density slice in the x-y plane through the center of the
      simulation of filament 2 in log-space. Only the filament gas is
      plotted ($\rho>10^{-20}\text{\gccm}$). The upper panel shows the
      filament at the beginning which we define to be 1 Myr before the
      final state. The lower panel shows the final state that is compared
      with the observations. The timestep of the middle panel is exactly in
      between the two. The contours show equi-density levels normalized
      on the maximum density of the respective timestep and increase
      linearly in log-space to always show five contours. As one can see,
      the form of the emerging density clump changes from prolate in the
      linear evolution phase to becoming more and more roundish in the
      non-linear phase, a behavior which was also seen in SPH-simulations
      executed by \citet{inutsuka1992}.}
      \label{fig:filamentcontour}
   \end{figure*}

   The numerical simulations were executed with the RAMSES code
   \citep{teyssier2002}. The code is capable of solving the discretised
   Euler equations in their conservative form on an Cartesian grid
   in 1D, 2D and 3D with a second-order Godunov scheme. For our simulation
   we used the MUSCL scheme (Monotonic Upstream-Centred Scheme for
   Conservation Laws, \citet{vanLeer1976}) in combination with the
   HLLC-Solver \citep{toro1994} and the multidimensional MC slope
   limiter \citep{vanLeer1979} was applied in order to achieve a
   total variation diminishing scheme. The gravity is solved using the
   built-in multigrid solver.

   We place the filament axis in the x-direction of a 3D box and use periodic
   boundary conditions in this dimension in order to simulate an infinite
   filament. We set the boxsize to twice the dominant perturbation
   length in order to resolve it correctly and to stay close to the
   observations. The boundaries in perpendicular directions
   of the filament are set as outflow condition. The potential of the ghost
   cells has to be set to zero in order to not introduce a
   gravitational focus to the box center due to the mixed boundary conditions.

   The filament gas is set to be isothermal with a temperature of 10 K and
   a molecular weight of $\mu=2.36$. The external gas surrounding the
   filament is fixed to be isobaric at all times and in pressure equilibrium
   with the boundary pressure of the filament:
   \begin{equation}
      P_b = \rho_b c_s^2
      \label{eq:boundarypressure}
   \end{equation}
   In order to minimize the effect of accretion we set the external
   density to a very low value of $10^{-4}$ times the filament boundary
   density. Real physical accretion would affect the growth of
   perturbations, but to quantify this is above the scope of this
   paper.

   As the boxsize is larger than the filament diameter, we employ
   adaptive mesh refinement (AMR), which allows us to keep the resolution
   low in the low-density external gas. We enforce a refinement of the
   central region to give us an effective resolution of $256^3$ for the
   dense filament gas, which is enough to fulfill the Truelove criterion
   for the maximum occurring density within a factor of 16 at all times
   \citep{truelove1997} aside from the late core collapse in filament 2
   where we still fulfill the Truelove criterion within a factor of 8. In order
   to check for consistency and rule out a limitation by resolution we also repeat
   the simulations with half the resolution and look for mayor differences between
   the results.

   The density perturbation is set according to \autoref{eq:rhoperturbation}
   with a very small amplitude of 1\% of the initial density, as the velocity
   has to adjust to the perturbed state. The phase of the density perturbation
   is set at random, but as the box is periodic in the direction of the
   filament axis, it has no influence on the solution.

\section{Simulations}
\label{sec:simulations}

   \begin{figure}
      \includegraphics[width=\columnwidth]{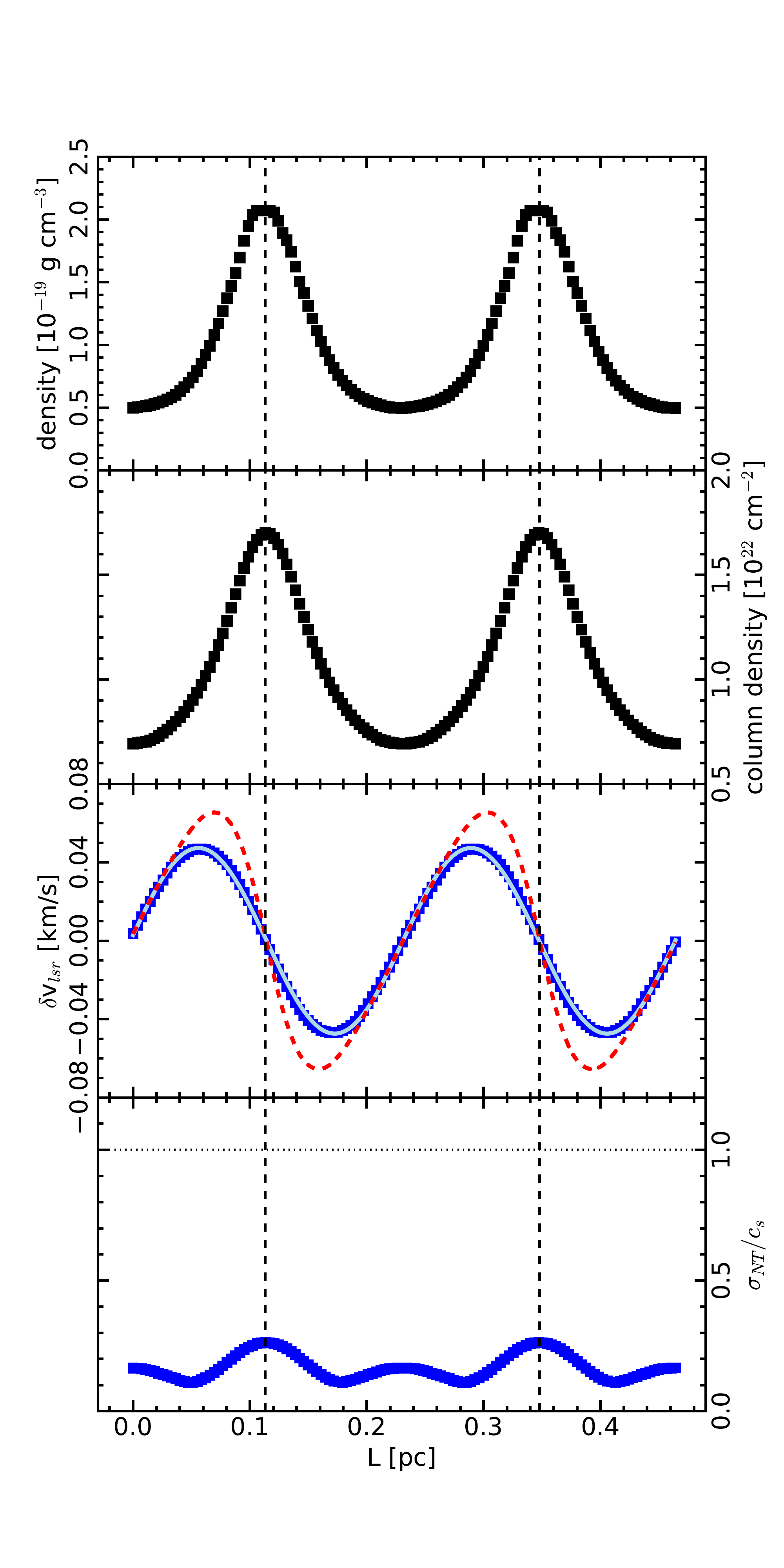}
      \caption{Results of the simulation for filament 1 in a line-of-sight
      projection of 57\degr along the axis of the filament.
      The upper panel shows the central volume density along the filament
      with two distinct cores also marked by the dashed, vertical lines.
      The density maximum and minimum matches the observation within a
      factor of two. The second panel shows the surface density summed up
      along the line-of-sight for a gas with a molecular weight of
      $\mu = 2.36$. The third panel shows the velocity centroid
      variation as blue squares. Our fit to the data is
      given by the light blue line and the true projected central axis
      velocity is shown by the dashed, red line. As one can see, the true
      central velocity pattern is hidden with the maxima being damped by the
      velocity structure inside the filament. The velocity centroid pattern
      shows nearly the same amplitude of $0.05 \text{ km s}^{-1}$ compared
      to the observed amplitude of $0.04 \text{ km s}^{-1}$. The lower panel
      shows the non-thermal velocity dispersion in units of the sound speed.
      The dispersion is subsonic throughout with a mean value of
      $\left\langle\sigma_{\text{nt}}\right\rangle/c_s \approx 0.17$ which
      is about a third of what is observed in filament 1
      ($\left\langle\sigma_{\text{nt}}\right\rangle/c_s = 0.57 \pm 0.15$
      \citep{hacar2011}).}
      \label{fig:filament1lineofsight}
   \end{figure}

   \begin{figure}
      \includegraphics[width=\columnwidth]{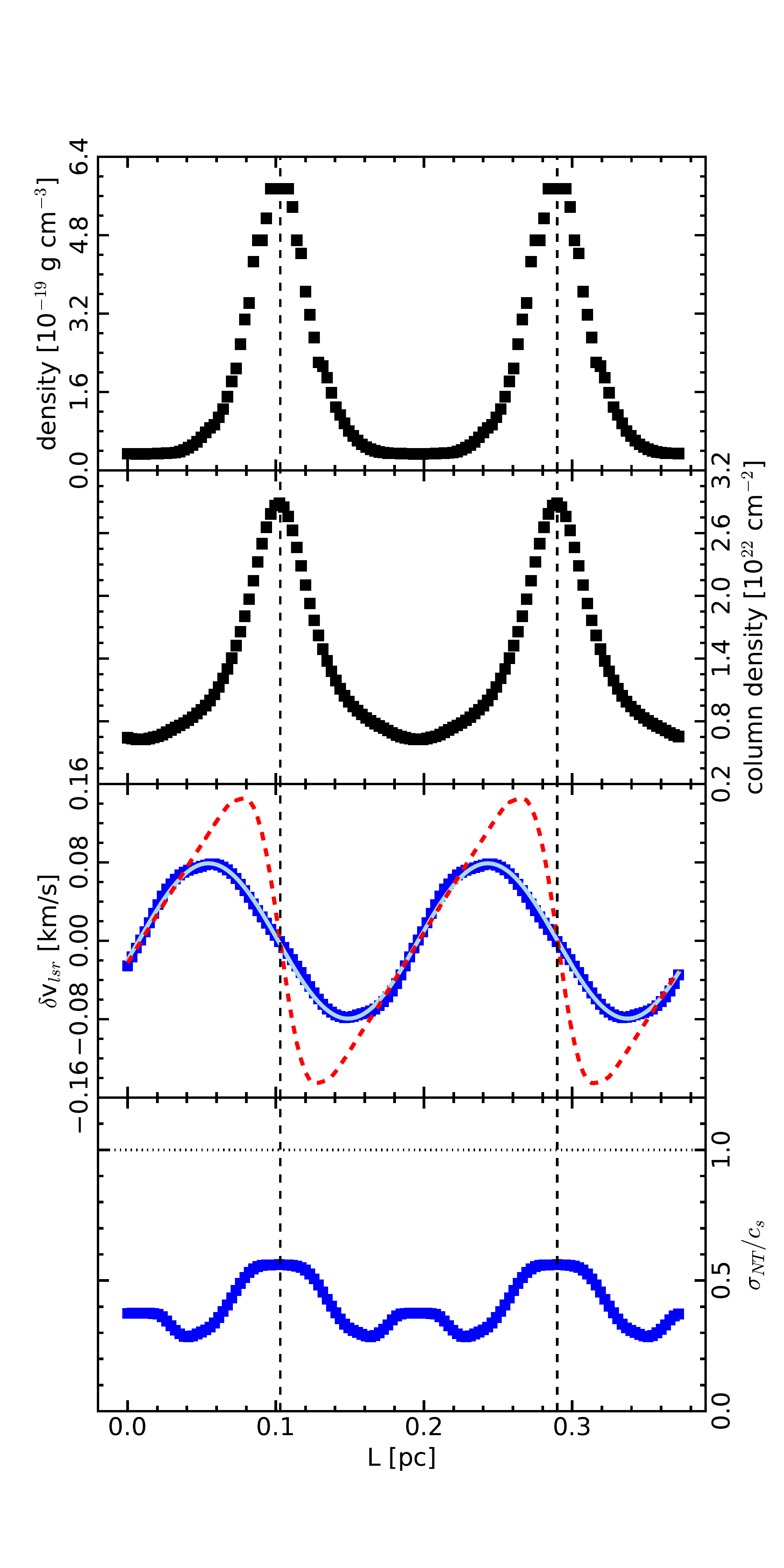}
      \caption{As in \autoref{fig:filament1lineofsight} but for filament 2 in
      a line-of-sight projection of 66\degr. Again, the upper panel shows
      that the volume density matches the observation within a factor of two.
      The cores are more pronounced than for filament 1 and the volume density
      does not follow a sinusoidal pattern along the filament but shows a
      flat decreased density between the cores. This feature is harder to see
      in the surface density in the second panel. The amplitude of the
      velocity centroid variation is somewhat higher than the fit of the
      observations, $0.08 \text{ km s}^{-1}$ compared to $0.04 \text{ km s}^{-1}$,
      but the original observed velocity data of filament 2 does not follow a well
      defined sinusoidal curve \citep{hacar2011}. The true central velocity
      structure is even more damped than in filament 1. It shows more pronounced
      slopes which indicates that matter from a broader region is pulled onto
      the density maxima. The non-thermal velocity dispersion in the lower panel
      is about twice that of the simulation of filament 1 with a mean value of
      $\left\langle\sigma_{\text{nt}}\right\rangle/c_s \approx 0.40$ which is
      around a factor of 1.5 lower compared to the observed value of
      $\left\langle\sigma_{\text{nt}}\right\rangle/c_s = 0.63 \pm 0.16$.}
      \label{fig:filament2lineofsight}
   \end{figure}

\subsection{Comparison with the observation}
\label{sec:observations}

   We let the simulations run until the maximum density matches the observed
   central core density. An example of a simulation can be seen in
   \autoref{fig:filamentcontour} where we plot a density slice through the center
   of the filament with overlaying density contours. The timescale is restricted by
   the lifetime of a molecular cloud, which is estimated to be of the order of
   a few Myr, especially for Taurus \citep{palla2000,hartmann2001,white2001}.
   It is therefore reasonable to assume 1 Myr as typical timescale for filament
   fragmentation. This timescale also
   coincides with the average lifetime of a core with typical densities of
   $10^4 \text{ cm}^{-3}$ \citep{lee1999,jessop2000,nakamura2005},
   which serves as a good approximation for the total starless core phase.
   In our case, as the evolution should be effectively self-similar in the
   linear phase, the evolution timescale is then a function of the initial
   amplitude of the perturbation. Starting with a very small initial perturbation
   and running the simulation until the observed state is reached, allows us to
   actually determine the most likely initial amplitude as the perturbations
   strength reached one Myr before the finale state.

   In order to compare our simulation result with the observations,
   we assume an inclination as described in \autoref{sec:prediction}. We then use
   this inclination to determine the line-of-sight velocity distribution for
   each spatial pixel. In order to get the centroid velocity, we treat every
   volume element of the computational grid as an emitter of a discrete
   line-of-sight velocity value. These are converted to Gaussian line profiles
   with a dispersion of $\sigma = 0.0526 \text{ km s}^{-1}$, corresponding to the
   thermal linewidth of C$^{18}$O, and are weighted with the respective density.
   The line profiles are binned into histograms with a bin width of
   $0.05 \text{ km s}^{-1}$ in order to get a complete line emission for each
   observed spatial pixel and we measure the velocity centroid by fitting a
   Gaussian to the line, as an observer would do.

   The result for our converged models is shown in
   \autoref{fig:filament1lineofsight} for filament 1 and
   \autoref{fig:filament2lineofsight} for filament 2 together with the
   central volume density, the column density in the line-of-sight and
   the non-thermal velocity dispersion of the Gaussian fit.
   Filament 1 has an unprojected wavelength of $0.43 \text{ pc}$, an
   inclination angle of 57\degr and an initial line-mass of $f_\text{cyl}=0.51$.
   Filament 2 has an unprojected wavelength of $0.46 \text{ pc}$, an
   inclination angle of 66\degr and an initial line-mass of $f_\text{cyl}=0.43$.
   The projected core distances of $0.23 \text{ pc}$ and $0.19 \text{ pc}$
   respectively match the observed spacing and the maximum central volume
   densities of $2.07 \cdot 10^{-19} \text{\gccm}$ and
   $5.74 \cdot 10^{-19} \text{\gccm}$ also agree with the
   observations. The minimum of the central volume densities of
   $4.98 \cdot 10^{-20} \text{\gccm}$ and
   $3.37 \cdot 10^{-20} \text{\gccm}$ match
   the observed densities between the cores within a factor of two,
   which is the estimated error of measurement of \citet{hacar2011}.

   The spacial distribution of the line-of-sight centroid velocities is in
   good agreement with the observations. Especially filament 1 where the
   observed velocity data shows a sinusoidal curve with an amplitude of
   $0.04 \text{ km s}^{-1}$ is well matched. Our fit, given by the solid
   light blue line, has an amplitude of $0.05 \text{ km s}^{-1}$ but this
   is still within the measurement error of $0.01 \text{ km s}^{-1}$. \citet{hacar2011}
   also fitted an amplitude of $0.04 \text{ km s}^{-1}$ to the line-of-sight
   centroid velocity distribution of filament 2 but the observed velocity
   structure is ambiguous. The observed centroid velocity of filament 2
   varies up to a value of $0.1 \text{ km s}^{-1}$ and does not follow a
   sinusoidal pattern very well. Contrarily to the observation, we see a
   sinusoidal pattern in the centroid velocity variation. We find an amplitude
   about twice as large as the observed fit, namely $0.08 \text{ km s}^{-1}$.
   Why the observed centroid velocity does not follow a clear pattern is not
   clear and could be due to more complex motions inside the filament.
   The velocity patterns of both filaments show a clear $\lambda/4$
   shift in phase compared to the density perturbation, as they do in the
   observations, which is an indicator of the core-forming motions
   \citep{gehman1996}.

   The dashed, red lines in \autoref{fig:filament1lineofsight} and
   \autoref{fig:filament2lineofsight} show the true,
   inclination corrected, central axis velocity in longitude direction of the
   filament. Due to the central axis being the densest component, we would have
   expected that this velocity agrees with the projected line-of-sight velocity
   pattern as we only see weak radial infall motions in the majority of
   the filament. However, contrarily to our expectation, the true motion differs
   significantly from the line-of-sight velocity structure. Its maxima are large
   with a difference that is almost a factor of 2 in filament 2. In addition,
   the location of the maxima is shifted towards the clumps. In the strong
   non-linear case of filament 2 the true velocity pattern cannot be modeled
   well by a sinusoidal pattern. Interestingly, however, the projected velocity
   structure still resembles a sinusoidal pattern well.

   The lowest panel shows the non-thermal velocity dispersion in the
   line of sight through the centre of the filament. It is calculated by
   measuring the FWHM, here denoted by $\Delta v$, of the Gaussian fit to
   the line-of-sight velocity data and by using the fact that the
   contribution of the thermal and non-thermal gas motions to the linewidth
   add in quadrature \citep[e.g.][]{myers1983}:
   \begin{equation}
       \sigma_\text{nt} = \sqrt{\frac{\Delta v^2}{8 \ln 2}-\frac{k_B T}{m}}
       \label{eq:nonthermalsigma}
   \end{equation}
   where $m$ is the mass of the observed molecule, in this case C$^{18}$O.
   We find that the gas is very subsonic throughout the filaments, with
   variations due to higher velocity dispersions inside and between the
   cores. We find that the higher velocity dispersion inside the cores is
   not only due to the radial collapse of the filament but is dominated by the
   infall of gas coming from the bulk of the filament. It is interesting to
   note that, although we do not model subsonic turbulence, the ordered motions
   inside the filament already lead to a considerable amount of non-thermal
   velocity dispersion. The mean of the dispersion in units of the sound speed
   is about $\left\langle\sigma_{\text{nt}}\right\rangle/c_s \approx 0.17$ for
   filament 1 and
   $\left\langle\sigma_{\text{nt}}\right\rangle/c_s \approx 0.40$ for
   filament 2. Comparing the modeled velocity dispersion to the observed
   values of $\left\langle\sigma_{\text{nt}}\right\rangle/c_s = 0.57 \pm 0.15$
   and $\left\langle\sigma_{\text{nt}}\right\rangle/c_s = 0.63 \pm 0.16$
   respecitvely and taking into account that velocity dispersions add in
   quadrature it becomes clear that in the case of the observations subsonic
   turbulence still dominates the non-thermal component of the velocity
   dispersion even with strong underlying ordered motion.

   In the end, we find that the initial, unperturbed central density of the
   filaments was $8.2 \cdot 10^{-20} \text{\gccm}$ for filament 1 and
   $6.1 \cdot 10^{-20} \text{\gccm}$ for filament 2. Both lie considerably
   below the mean value of the maximum and minimum observed density of about
   $1.2 \cdot 10^{-19} \text{\gccm}$ for filament 1 and
   $2.9 \cdot 10^{-19} \text{\gccm}$ for filament 2. This signifies
   that there is a considerable amount of asymmetrical evolution where
   more mass is transferred to the density maxima than is taken from
   the density minima. This leads to an enhancement of the maximum density
   while leading to a slower density decrease between the cores. This effect
   is stronger in filament 2 than in filament 1 as can be seen from the bigger
   discrepancy of the initial central density and the observed density mean.

\subsection{Dynamical evolution}
\label{sec:evolution}

\begin{figure}
      \includegraphics[width=\columnwidth]{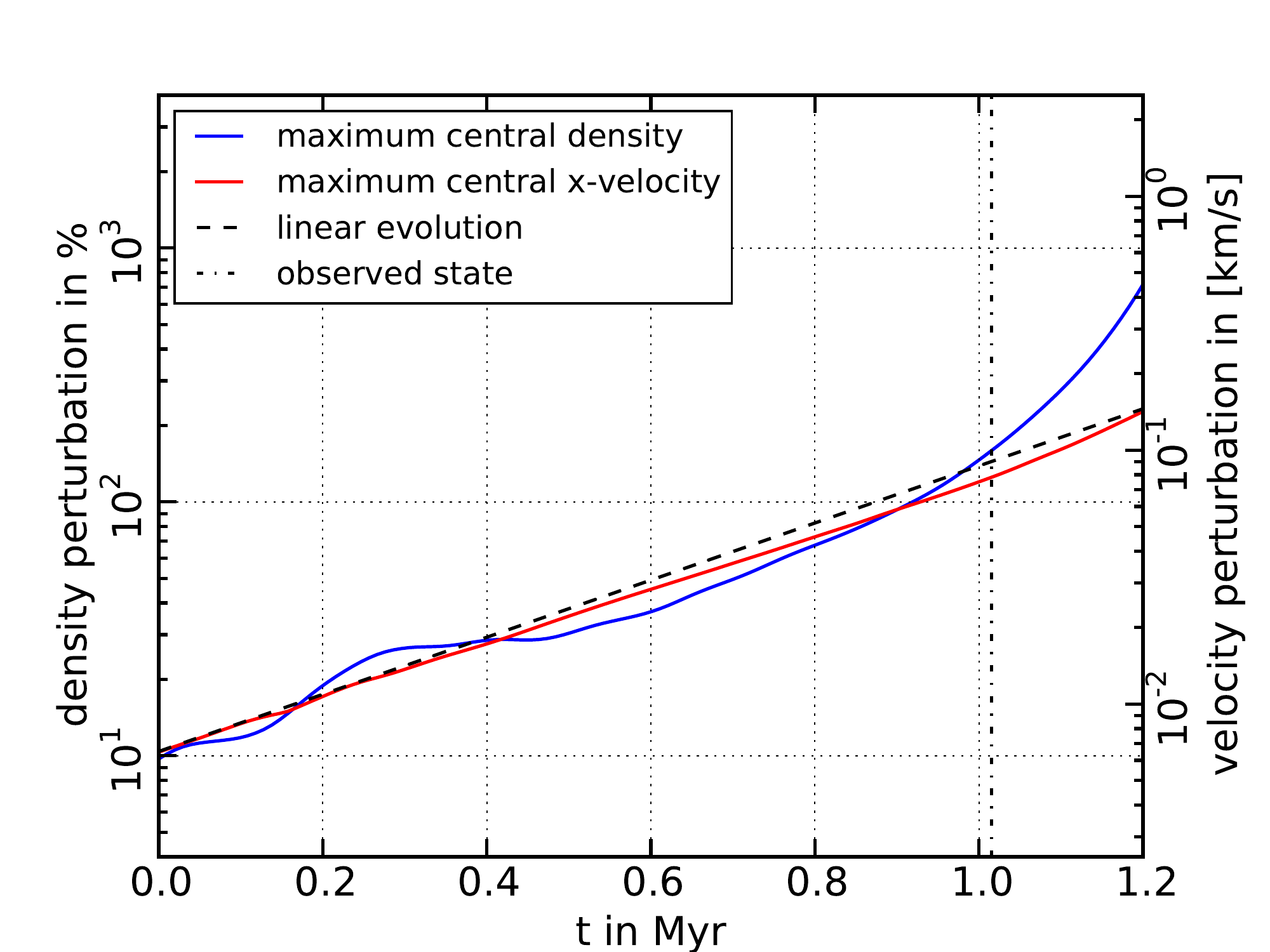}
      \caption{Evolution of the maximum central density (blue, left axis) and
      x-velocity (red, right axis) of filament 1. Both follow the analytic
      prediction until the density reaches a perturbation strength of about
      100\% where a non-linear evolution in density sets in.
      The dashed-dotted vertical line gives the point in time of the
      observation.}
      \label{fig:filament1evolution}
   \end{figure}

\begin{figure}
      \includegraphics[width=\columnwidth]{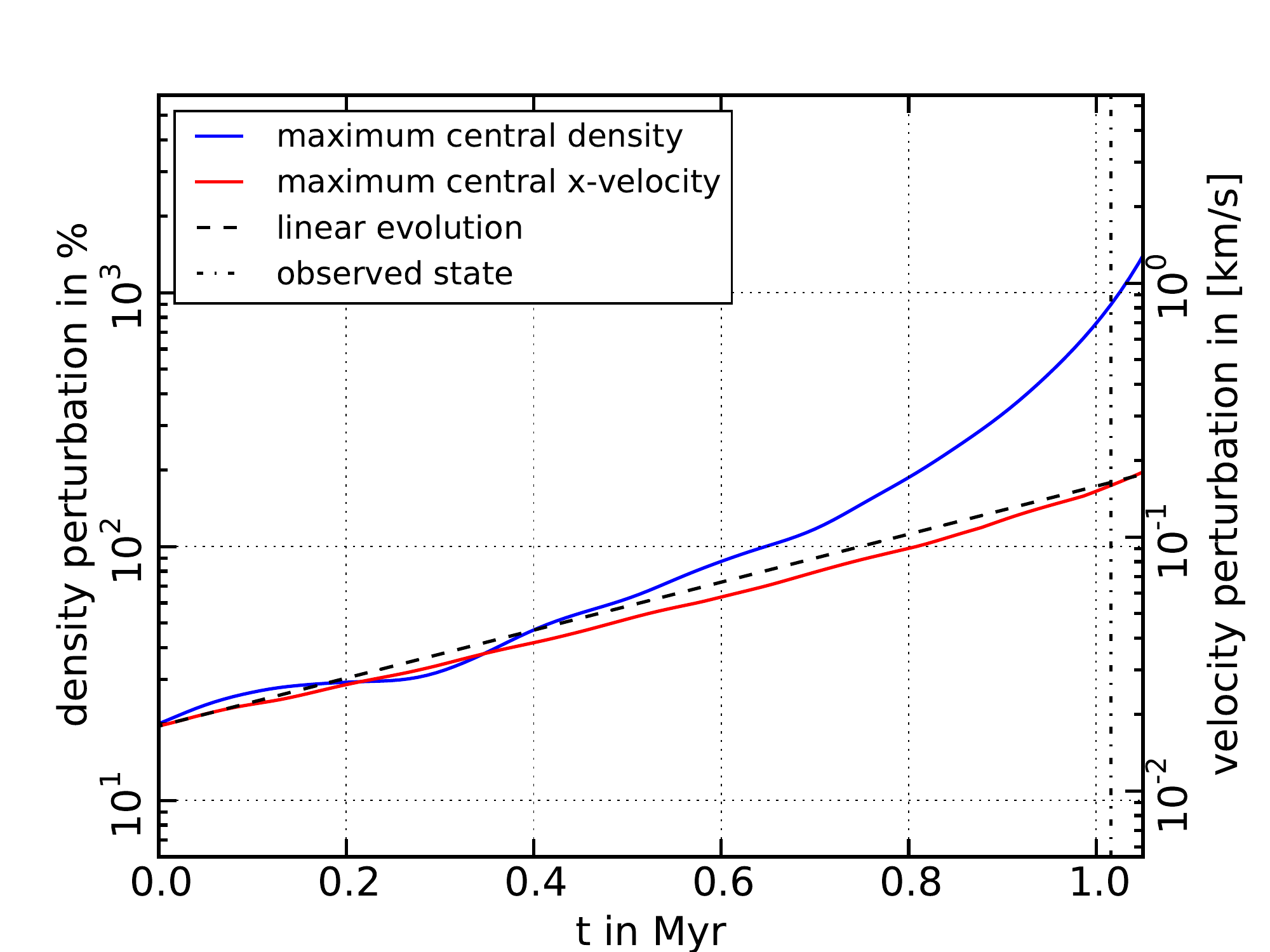}
      \caption{As in \autoref{fig:filament1evolution}, but for filament 2.
      Filament 2 shows the same evolutionary characteristics as filament 1,
      but in order to fulfill the time constraint of about 1 Myr, we need an
      initial density perturbation of around 20\%. One can see that the
      observed state lies far deeper in the non-linear regime than for
      filament 1.}
      \label{fig:filament2evolution}
   \end{figure}

   In order to understand the asymmetrical evolution of the maximum and
   minimum density, we take a closer look at the time evolution of the
   perturbed quantities. Each perturbation of every variable should follow
   \autoref{eq:perturbation}. Thus, they should follow a linear
   evolution in log-space. This is shown for the central maximum density
   and central maximum velocity in the filament axis in
   \autoref{fig:filament1evolution} for filament 1 and in
   \autoref{fig:filament2evolution} for filament 2. In order to stay close
   to a reasonable limit of 1 Myr (see previous subsection), we start the
   evolution of filament 1 at about 10\% perturbation strength and of filament
   2 at about 20\%. Both the maximum density and maximum velocity follow the
   linear prediction for the majority of the evolution with the density showing
   some oscillation around the linear prediction. In the late phase of the
   evolution, both filaments show a clear non-linear growth of the maximum
   density. It is worth noting that this non-linear phase is not short in the
   sense that all the mass of the filament collapses into a core in a fast
   period of time, but it is a smooth process, which can easily take up to
   0.4 Myr or nearly half of the whole formation time of a core and still
   leaves diffuse gas to make up the filament. Also note that the velocity
   does not follow the same non-linear evolution initially. Only when we let
   the simulations run further, we see that the velocity follows with a delay
   of about 0.3 Myr. At that point in time, the cores are in a much more
   developed state and the simulations brake down due to resolution issues.

   In order to have a deeper look into the non-linear phase we look at the
   line-mass in the slice of the maximum density. The line-mass of a
   pressure-bound filament is given by:
   \begin{equation}
      \frac{M}{L} = \int^R_0 2\pi r \rho(r) dr
   \end{equation}
   Inserting \autoref{eq:rhoperturbation}, one can show, as the perturbation
   is independent of radius and the integration just gives the initial
   line-mass, that the time evolution is given by:
   \begin{equation}
      \left(\frac{M}{L}\right)(t) =
      \left(\frac{M}{L}\right)_0[1+\epsilon\exp(ikz-i\omega t)]
      \label{eq:fcylperturbation}
   \end{equation}
   This indicates that, not only is there a redistribution of mass inside the
   filament, but that the line-mass also follows exactly the same evolution as
   the density.

\begin{figure}
      \includegraphics[width=\columnwidth]{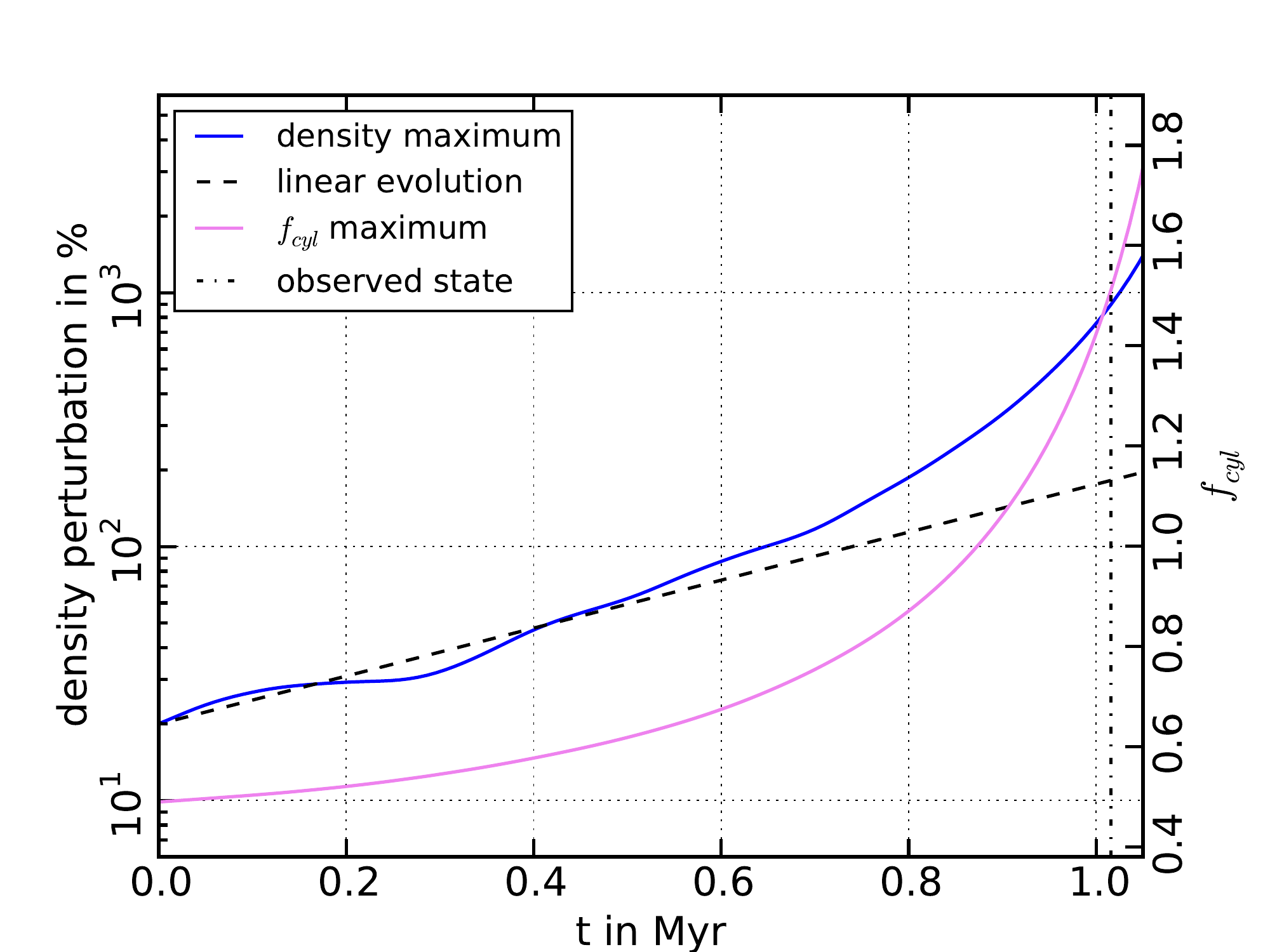}
      \caption{Evolution of the maximum density perturbation in log-space
      (blue, left axis, logarithmic scale) of filament 2 together with the
      evolution of $f_\text{cyl}$ (violet, right axis, linear scale) in the
      maximum density filament slice. One can see that the non-linear
      evolution of the maximum density is tied to the fact that the line-mass
      enters the critical regime ($f_\text{cyl}>1$). This means the maximum
      density evolution is driven primarily by the radial collapse of the
      filament in this section.}
      \label{fig:fcylevolution}
   \end{figure}

   We plot the time evolution of the maximum density together with its
   respective line-mass ratio $f_\text{cyl}$ of filament 2 in
   \autoref{fig:fcylevolution}. The scale of $f_\text{cyl}$ is linear in order
   to see when the line-mass becomes supercritical. One can clearly see that
   $f_\text{cyl}$ exceeds 1.0 after about 0.85 Myr. This coincides more or less
   with the maximum central density entering the non-linear part of its evolution
   and dominates the late evolution. This indicates that the reason for the
   non-linear evolution of the maximum central density is that the region
   containing the cores are supercritical in the line-mass. This leads to a
   radial collapse of the filament at the position of the core as it cannot
   sustain hydrodynamical equilibrium which in turn enhances the central density.

\begin{figure}
      \includegraphics[width=\columnwidth]{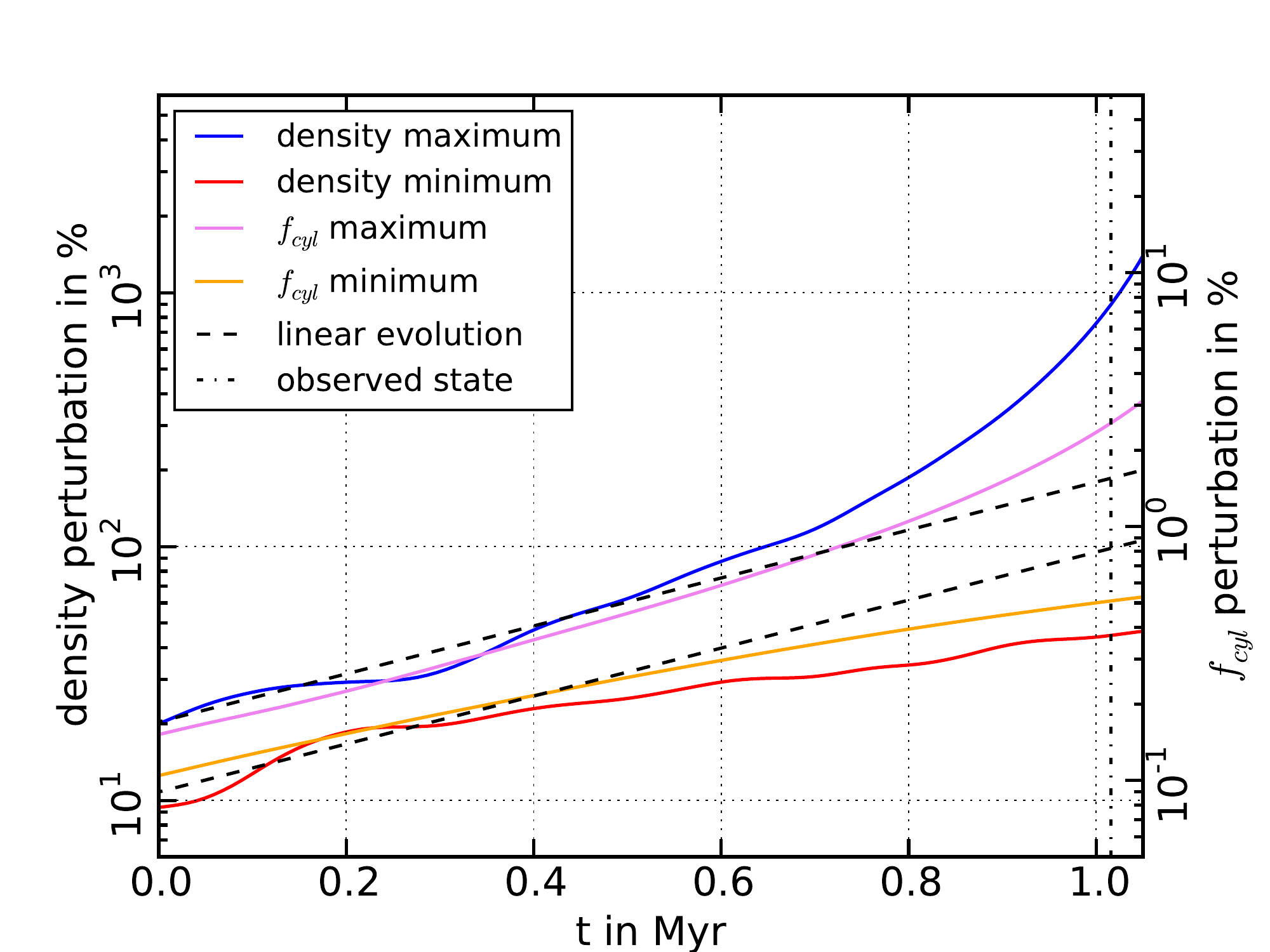}
      \caption{Evolution of the maximum and absolute minimum value of the
      density perturbation (left axis) of filament 2 together with the
      evolution of the absolute value of the perturbation in $f_\text{cyl}$
      (right axis) in the respective filament slice. The minimum density
      perturbation evolves slower as predicted, growing even slower as time
      progresses. This means that the cores are not feeded symmetrically
      by the density minima, but accrete mass from over the whole filament.}
      \label{fig:fcylminimum}
   \end{figure}

   This effect can account for the majority of the asymmetry but we also
   observe a reduced growth of the perturbation of the density minimum. In
   \autoref{fig:fcylminimum} we plot the evolution of the maximum and the
   minimum density as well as the evolution of the maximum and minimum
   line-mass, both on a logarithmic scale to show the linear evolution.
   Note that, not only does the density follow the linear prediction
   but also the line-mass as predicted by \autoref{eq:fcylperturbation}.
   One can also see the non-linear evolution of the maximum density in the
   line-mass maximum but it does not grow as fast as the density. As radial
   collapse of the filament would not lead to a difference in line-mass,
   this indicates that the density growth is not fed by radial collapse alone.

   An interesting effect is that the evolution of the minimum density,
   as well as the minimum line-mass, flattens as the perturbation of the
   maximum becomes non-linear. This implies that the mass is not
   redistributed from the minimum to the maximum anymore
   but that the cores accrete mass from
   all over the bulk of the filament, which can be seen in the central
   volume density of filament 2 in \autoref{fig:filament2lineofsight}.
   The area between the cores shows a flat decreased density,
   with the cores being very pronounced and peaked, indicating that the
   cores also pull in mass from the filament axis. This means that as we
   approach the supercritical state of the cores, the accretion changes
   from a simple linear enhancement and reduction of the maximum and minimum
   respectively to a radial, spherical symmetric accretion onto the cores.
   This effect has also been shown and studied by \citet{inutsuka1992} in
   SPH-simulations where they see the same behavior in the non-linear phase.

   The change from linear to non-linear evolution also leads to a
   difference in core morphology and has an important implication for
   observed cores. This change can be seen in \autoref{fig:filamentcontour}
   where the density contours show the cores initially forming to have a
   prolate shape, as matter is only transferred from the minimum to the maximum
   density. This changes drastically in the non-linear phase at the end of the
   simulation where the dense cores clearly show a generally round form in the
   density distribution. This was also seen in the simulations of
   \citet{inutsuka1992} where the cores approached a near spherical form for
   late times in the non-linear phase. If one were to observe a dense core that
   displays an elongated, prolate form, it could be a strong hint that it is
   situated in the linear evolutionary phase. Indeed, observed cores typically
   show an elongated form \citep{benson1989} which is consistent with the fact
   that cores spend the bigger part of their lifetime in the linear evolutionary
   phase. However, it is obvious that there is an observational bias to detect
   dense cores which are likely to be in the non-linear phase already as they
   would not easily stand out of the filament gas otherwise. For instance,
   at the beginning of the non-linear phase of the simulation of filament 2
   the cores only have an over-density of a factor of about two with respect
   to the filament gas.

   The factor of two difference in the initial perturbation of the density
   maximum and the density minimum in \autoref{fig:fcylminimum} stems from the
   fact that we start the simulation at an even earlier time with a very small
   perturbation strength in order to let the filament adjust to the perturbation.
   Although it is a relatively small difference, it highlights the fact that
   the density minimum already evolved slower than the density maximum. Starting
   with different values for the initial perturbation strength shows that the
   factor of two is robust and seems to be hard to avoid. Nevertheless, it is
   of relatively minor impact, since a factor two is also the inherent error
   in the observation and would not lead to an offset of the mean density as
   initial condition.

\subsection{External pressure}
\label{sec:externalpressure}

   From \autoref{eq:rhoboundary}, it is possible to determine the boundary
   density of the filaments. This value can then be used in
   \autoref{eq:boundarypressure} to get an estimate of the pressure of the
   surrounding material. Using our derived values of
   $\rho_0=8.18 \cdot 10^{-20} \text{\gccm}$ and $f_\text{cyl}=0.51$ for
   filament 1 and $\rho_0=6.12 \cdot 10^{-20} \text{\gccm}$ and
   $f_\text{cyl}=0.43$ for filament 2, both filaments give nearly the exact same 
   value for the external pressure of
   $P_\text{ext}/k_B = 5.01 \cdot 10^4 \text{\Kccm}$ and
   $P_\text{ext}/k_B = 5.08 \cdot 10^4 \text{\Kccm}$ where $k_B$ is the
   Boltzmann constant. The consistency of both values
   could be a coincidence, but is reassuring and shows the power of the
   pressure truncated filament method to predict the environmental
   pressure.

   However, the source of the external pressure is not clear and is open to
   debate. It is larger than the usually assumed total (thermal + turbulent)
   gas pressure of the interstellar medium, which is estimated to be of the
   order of $P_\text{ISM}/k_B \approx 10^3-10^4 \text{\Kccm}$, e.g. by
   \citet{bertoldi1992}, although they also estimate the gravitational
   pressure of the weight of the overlying material of different molecular
   clouds on cores and get values in a range of
   $P_G/k_B \approx 10^4-10^5 \text{\Kccm}$. While this pressure is more
   important in quiescent clouds and not necessarily as high for turbulent
   clouds, gravitational pressure can add to the overall external pressure.

   In addition, warm ionized gas can also lead to a supportive pressure, as has
   been shown for the Pipe Nebula \citep{gritschneder2012}. They also
   determine pressure values in the range of
   $P_\text{ion}/k_B \approx 10^4-10^5 \text{\Kccm}$. A possible source
   for ionizing radiation is AB Aurigae, which has been shown to heat the
   diffuse gas component while having nearly no influence on the dense gas
   \citep{duvert1986,ladd1991}. However, for the estimated environmental
   densities of order $5 \cdot 10^2 \text{ cm}^{-3}$ we would require a
   temperature of around 100 K which is not observed in the external medium.

   Another source of external pressure is turbulent
   ram pressure $P_\text{ram} = \rho\sigma_\text{turb}^2$ where
   $\sigma_\text{turb}$ is calculated according to \autoref{eq:nonthermalsigma}.
   The relevant velocity dispersion $\sigma_\text{turb}$ for an external ram
   pressure is not the intrinsic velocity dispersion of the filaments but
   that of the whole region. The observations of \citet{hacar2011} show that
   the C$^{18}$O gas is measured over the range of
   $\Delta v = 1.2 \text{ km s}^{-1}$. Assuming an external density of
   $\rho_b \sim 5 \cdot 10^2 \text{ cm}^{-3}$ leads to an external pressure
   of about $P_\text{ram}/k_B \approx 4 \cdot 10^4 \text{\Kccm}$
   which is in excellent agreement with our external pressure estimate.

   Although the source of the external pressure still remains unclear, a good
   candidate is therefore turbulent ram pressure. The theoretical model does not
   allow for a big leeway in the external pressure estimate and further
   observations are required to confirm this conjecture and determine the origin
   of the external pressure.

\section{Discussion and conclusions}

   Even though we can reproduce the observational data well, some caveats remain.
   First and foremost, we simulate an infinite filament by
   using periodic boundary conditions. This is not a perfect representation of reality
   and our choice of boundary prevents the global collapse and edge-effects expected in
   finite filaments. A future study must include effects of a non-periodic boundary.
   It was shown by \citet{clarke2015} using
   semi-analytical methods and simulations that the global timescale for collapse of
   filaments with greater aspect ratios than $A>2$ is:
   \begin{equation}
      t_\text{col}= \frac{0.49+0.25A}{\sqrt{G\frac{(M/L)}{\pi R^2}}}
   \end{equation}
   Using this formula, we find that the timescales for global collapse of
   our filaments are around 1.5 Myr. Thus, although fragmentation is
   physically possible, the filament will also
   change its form on a Myr scale. This will definitely influence
   fragmentation length- and timescales and if it is not possible to find
   other stabilizing processes, e.g. rotation or magnetic fields, this
   effect has to be taken into account.

   Furthermore, although we can explain the observed densities, some
   differences to the observations remain. Most notably,
   the cores in \autoref{fig:filament1lineofsight} and
   \autoref{fig:filament2lineofsight} are more peaked than the observed
   cores. Also the observed filaments have a diameter of about 0.2 pc in
   contrast to our simulations where the filaments have a diameter of about
   0.14 pc. The reason for this could be that we do not include the intrinsic
   turbulence. \citet{hacar2011} do find that a considerable amount of
   sub-sonic turbulence of about 0.5 times the sound speed dominates the
   filaments. This will naturally lead to puffed-up cores and a wider filament
   itself. Turbulence could indeed also change the form of the filament
   and fragmentation length- and timescales and is a factor to take into
   account but as it does not change the line-mass that regulates the
   fragmentation the impact should be relatively small. Still, a detailed
   study on the effect of turbulence would be interesting in order to
   understand better whether the fragmentation presented here is still
   possible.

   Additionally, the observed cores are not symmetrical. For an idealized analysis we
   treat them as symmetric with a central density corresponding to their mean, but in
   reality their central density differs up to a factor of about three for filament 2.
   This could be either an effect of a clumpy or uneven initial mass distribution or
   of an asymmetrical evolution out of an evenly mass-distributed idealized filament.
   In an extended study, it should be possible to break their symmetrical evolution by
   introducing a power spectrum on the initial perturbation instead of only using the
   dominant wavelength.

   Nevertheless, we have shown that the fragmentation length scale can indeed be
   explained by subsonic gravitational fragmentation of the filament, assuming
   an idealized model. Together with a constraint on the
   inclination we can estimate line-of-sight centroid velocity variations and
   compare them to the observations.
   Our models give the most likely properties to be:
   \begin{description}
      \item[\textbf{Filament 1}]
   \end{description}
   \begin{itemize}
      \item an unprojected length of 0.95 pc
      \item an inclination of 57\degr
      \item a line-mass of $f_\text{cyl}=0.51$
      \item an initial central density of
            $\rho_0=8.18 \cdot 10^{-20} \text{\gccm}$
      \item an external pressure of $P_\text{ext}/k_B = 5.01 \cdot 10^4 \text{\Kccm}$
   \end{itemize}
   \begin{description}
      \item[\textbf{Filament 2}]
   \end{description}
   \begin{itemize}
      \item an unprojected length of 1.03 pc
      \item an inclination of 66\degr
      \item a line-mass of $f_\text{cyl}=0.43$
      \item an initial central density of
            $\rho_0=6.12 \cdot 10^{-20} \text{\gccm}$
      \item an external pressure of $P_\text{ext}/k_B = 5.08 \cdot 10^4 \text{\Kccm}$
   \end{itemize}
   Moreover, we demonstrated that cores can spend a considerable amount of their
   lifetime in a non-linear phase where their central density grows faster than a
   simple symmetrical mass transfer from the density minimum to the core. This is due
   to the fact that the densest regions of a filament exceed the line-mass where a
   radial hydrodynamic equilibrium is possible. This does not lead to an instant
   radial collapse, but the cores accrete matter radially, as well as from the whole
   filament. This makes them live long enough to be observed in the non-linear phase.

   Most importantly, the change from linear to non-linear evolution is indicated in a
   change of core morphology. While a symmetrical redistribution of material from the
   minimum to maximum density leads to a prolate form, the radial collapse in all
   directions of the cores in the non-linear phase makes the core roundish in
   appearance. The high density contrasts from core to filament gas which are only
   achieved in the non-linear evolution makes cores more likely to be observed in the
   non-linear phase.

\section*{Acknowledgements}

   We thank Matthias Gritschneder and the whole CAST group for helpful
   comments and discussions. We also thank the anonymous referee for improving
   the structure of the paper. SH wants to thank Alexander Beck for useful comments that
   clarified the presentation of the paper. AB and SH are supported by the priority
   programme 1573 "Physics of the Interstellar Medium" of the German Science Foundation
   and the Cluster of Excellence "Origin and Structure of the Universe"



\bibliographystyle{mnras}
\bibliography{L1517.bib}







\bsp	
\label{lastpage}
\end{document}